\documentclass[%
groupedaddress,
 amsmath,amssymb,
 aps,
floatfix,
]{revtex4-1}

\usepackage{graphicx}% Include figure files
\usepackage{graphics}% Include figure files
\usepackage{dcolumn}% Align table columns on decimal point
\usepackage{bm}% bold math
\usepackage{amsmath}
\usepackage{amssymb}
\usepackage{subfig}
\usepackage{float}

\usepackage{verbatim}
\usepackage{booktabs}
\usepackage{color}

\begin{document}

%\title{Hydrodynamic bound states in suspensions of microrollers: a dynamical system insight}%

\title{Hydrodynamically-bound states of a pair of microrollers: a dynamical system insight}

\author{Blaise Delmotte}
\email{blaise.delmotte@ladhyx.polytechnique.fr}
 \affiliation{LadHyX, UMR CNRS 7646, \'Ecole Polytechnique, 91128 Palaiseau CEDEX, France}

\date{\today}
\begin{abstract}
Recent work has identified persistent cluster states which were shown to be assembled and held together by hydrodynamic interactions alone [Driscoll  \textit{et al.} (2017) Nature Physics, 13(4), 375]. These states were seen in systems of colloidal microrollers; microrollers are colloidal particles which rotate about an axis parallel to the floor and generate strong, slowly decaying, advective flows. To understand these bound states, we study a simple, yet rich, model system of two microrollers. 
Here we show that pairs of microrollers can exhibit  hydrodynamic bound states whose nature depends on a dimensionless number, denoted $B$, that compares the relative strength of gravitational forces and external torques. 
Using a dynamical system framework, we characterize these various states in phase space and analyze the bifurcations of the system as $B$ varies.
In particular, we show that there is a critical value, $B^*$, above which active flows can beat gravity and lead to stable motile orbiting, or ``leapfrog", trajectories, reminiscent of the  self-assembled motile structures, called ``critters", observed by Driscoll  \textit{et al}. We identify the  conditions for the emergence of these trajectories and study their basin of attraction. 
This work shows that a wide variety of stable bound states can be obtained with only two particles.
  Our results aid in  understanding the mechanisms that lead to spontaneous self-assembly in hydrodynamic systems, such as microroller suspensions, as well as how to optimize these systems for particle transport. 
%In this paper we investigate the 
%Small particles rotating about an axis parallel to a no-slip boundary translate and generate strong, slowly decaying, advective flows.  Here we show that pairs of microrollers can exhibit various hydrodynamic bound states whose nature depends on a dimensionless number, denoted $B$, that compares the relative strength of gravitational forces and external torques. Using a dynamical system framework we characterize these various states in the phase space.  We analyze the bifurcation of the fixed points and limit cycles as $B$ varies. In particular, we show that there is a critical value of $B$ for the existence of stable motile orbiting trajectories, reminiscent of the "critters" observed by Driscoll \textit{et al.} \cite{Driscoll2017}. These results help us understand the hydrodynamic mechanisms that lead to spontaneous self-assembly and to envision particle transport in microroller suspensions.
\end{abstract}

\pacs{Valid PACS appear here}% PACS, the Physics and Astronomy
                    
\maketitle

\section{Introduction}
\begin{figure}
    \centering
    \includegraphics[width=0.9\columnwidth]{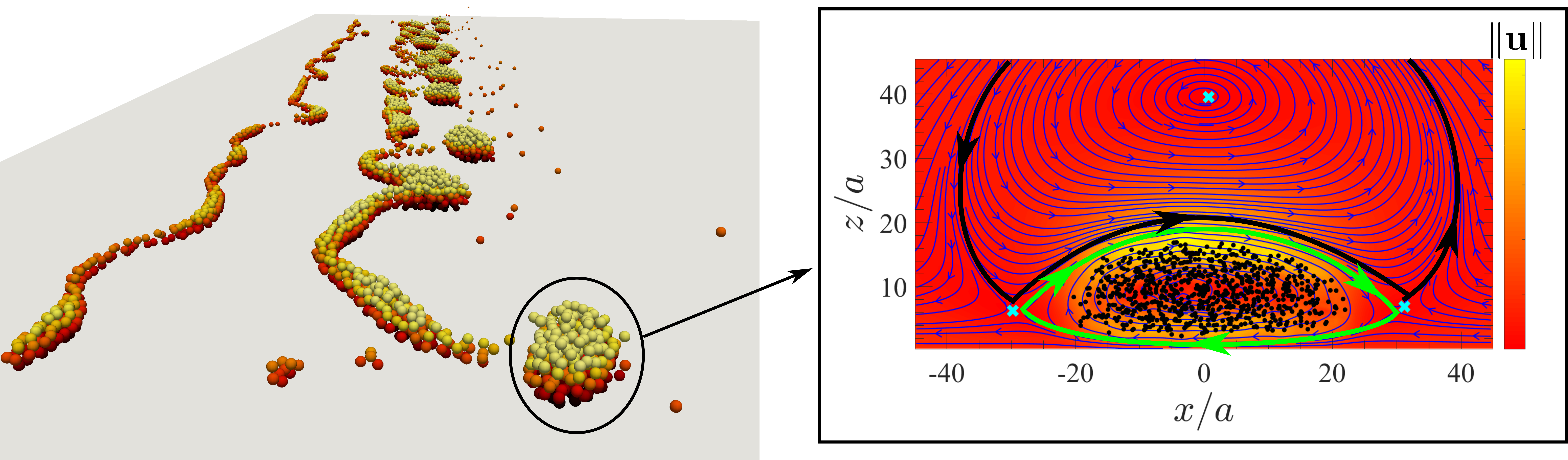}    \caption{Left: detachment of self-sustained motile clusters made of hundreds of microrollers, called critters. Particles are colored according to their translational speed. Right: cross-section of the flow field in the frame of a moving critter in the $xz$-plane. Black circles correspond the particles in the critter. Cyan crosses:  stagnation points of the flow. The green line with arrows delimits the recirculation zone where the particles perform a treadmilling motion. The black line with arrows indicates the counter-rotating recirculating flow above the critter. Colorbar: magnitude of the flow field.} 
    \label{fig:critters}
\end{figure}

Synchronization \cite{Lauga2009,Golestanian2011} and  collective  motion \cite{Vicsek2012,Marchetti2013,Bechinger2016}  are well documented in the literature on active and driven suspensions at low Reynolds number. The strong, slowly decaying, flow fields induced by the motion of small objects immersed in a viscous fluid play a predominant role in the emergence of coherent structures. Turbulent-like flows in bacteria and sperm suspensions \cite{Wu2000,Wensink2012,Dunkel2013,Creppy2015}, and phase synchronization between flagella \cite{Brumley2014} are illustrative examples in natural systems. 
Colloidal particles have been designed to mimic the behavior of natural systems at both the individual and collective level \cite{Fischer2011,Bruot2016}, and also to explore alternative ways to mix the surrounding fluid or to transport particles at low Reynolds number \cite{Snezhko2016,Aubret2017,Driscoll2018}. In particular, recent experiments and simulations have shown that suspensions of torque-driven particles above a floor could self-assemble into stable motile structures, called ``critters", that have no analog in natural systems \cite{Driscoll2017}, see Figure \ref{fig:critters}.
While it is clear that long-ranged hydrodynamics play a leading role in this phenomenon, the conditions for this spontaneous self-assembly must be identified.

The physics of torque-driven colloidal particles (called microrotors) have recently attracted significant attention \cite{Snezhko2016,Martinez2018,Driscoll2018}. Microrotors can be driven with an external rotating magnetic field \cite{Tierno2008,Driscoll2017}, or by using a Quincke-like instability under the action of an electric \cite{Bricard2013} or magnetic field \cite{Kokot2017}. Despite their apparent simplicity,  microrotors can lead to complex and interesting  dynamics at the collective level. Microrotors can be divided into two categories: (1) microspinners, which rotate in the absence of boundary or with an axis of rotation perpendicular to the interface, and (2) microrollers, which rotate about an axis parallel to the boundary (without necessarily touching it), and thus translate due to the asymmetric stress distribution on their surface.
 Previous studies have shown that microspinners can phase separate \cite{Yeo2015}, mix the surrounding medium \cite{Yeo2016} and arrange into large-scale rotating structures \cite{Snezhko2016,Kokot2017}. In proximity to the floor, suspensions of microrollers form traveling  waves \cite{Bricard2013,Delmotte2017} and shocks   \cite{Driscoll2017,Delmotte2017} that can destabilize into finger-like structures whose dense tips detach and self-assemble into stable motile clusters, called critters \cite{Driscoll2017} (cf. Fig. \ref{fig:critters}).
While the hydrodynamic mechanisms for the shock formation and fingering instability are well understood \cite{Delmotte2017,Delmotte2017b}, the conditions for the detachment of fingertips and their self-assembly into stable motile structures are not well identified. 
From a dynamical systems point of view, these critters could be viewed as an attractor whose existence and basin of attraction depend on the parameters of the system. 
Our goal here is to find the relevant parameters of the system to identify the conditions for the existence and stability of such hydrodynamic bound states. 

Hydrodynamic bound states of active particles near boundaries have been observed in various forms \cite{Lauga2009,Martinez2018}. At the individual level, sperm cells and bacteria swim in circles near boundaries \cite{Gadelha2010,DiLeonardo2011, Shum2015}, while biflagellate microorganisms and other model swimmers can be trapped \cite{ Spagnolie2012,Kantsler2013,Lushi2017} or perform periodic vertical motion \cite{Crowdy2010,Crowdy2011,Crowdy2011b,Davis2012}.
At the collective level, Martinez-Pedrero \textit{et al.} \cite{Martinez2018b} showed that the lateral flows generated by heavy microrollers can form horizontal one dimensional arrays of aligned particles.

\begin{figure}
    \centering
    \subfloat[][ ]{\includegraphics[width=0.4\columnwidth]{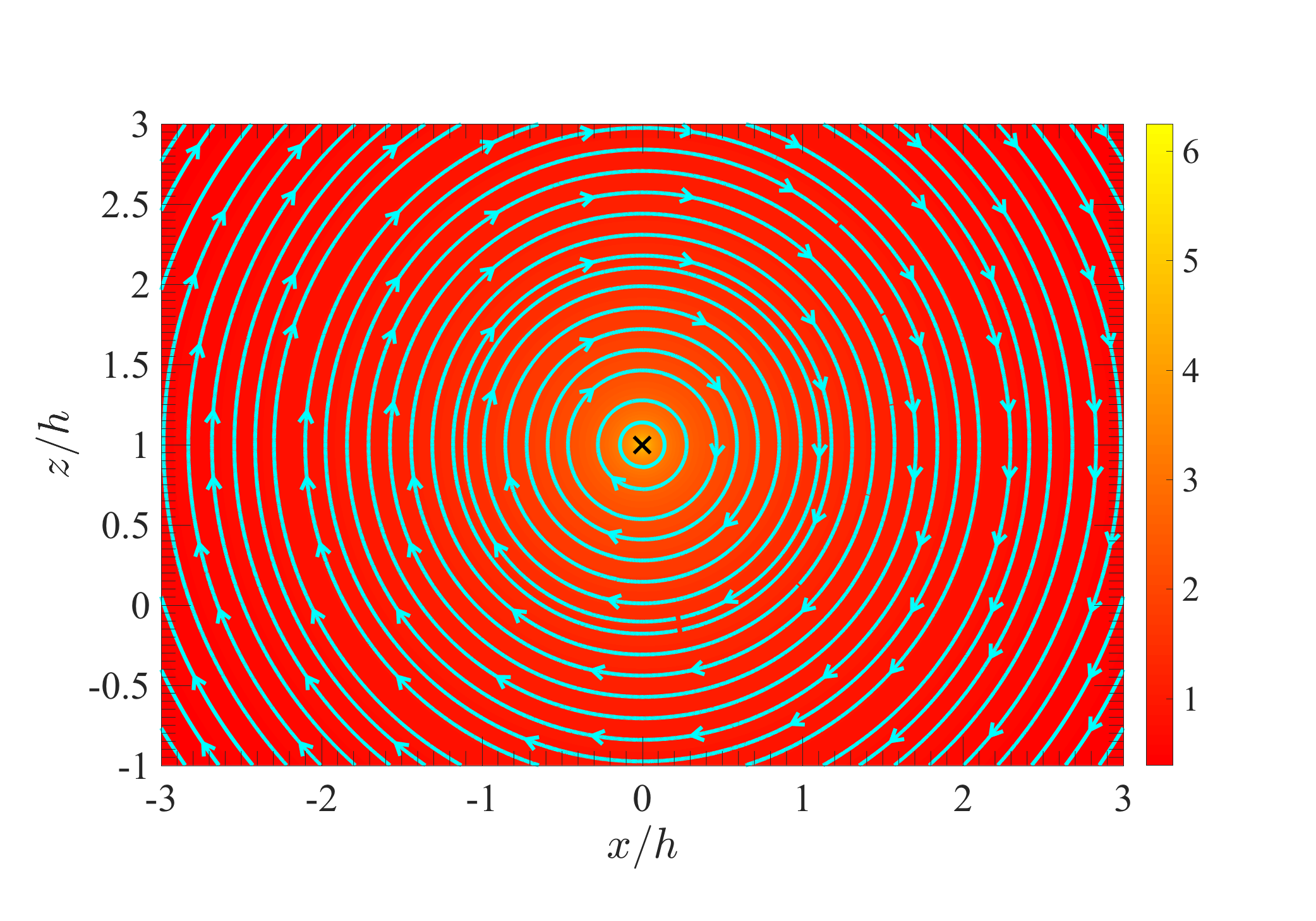}}
    \subfloat[][]{\includegraphics[width=0.4\columnwidth]{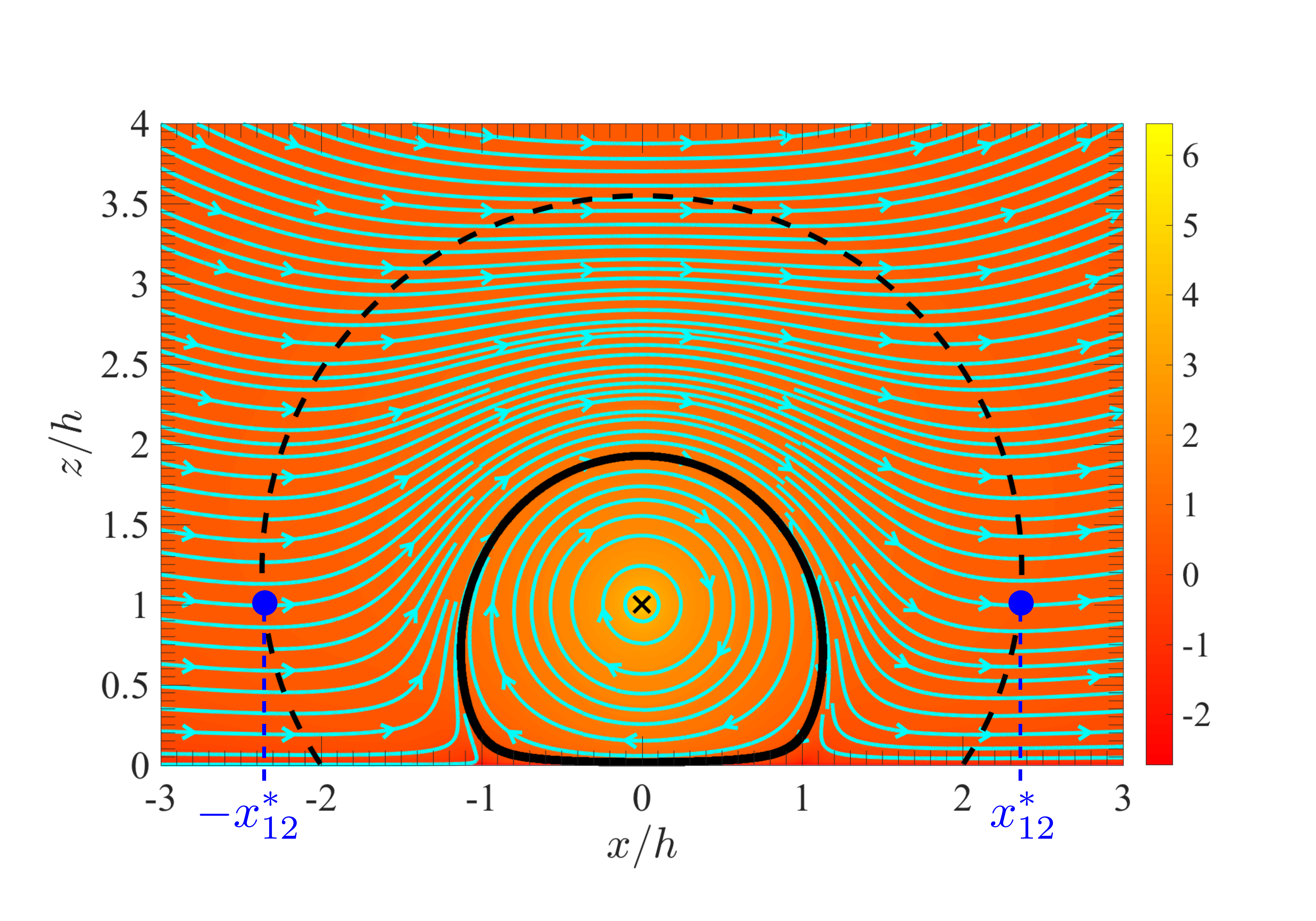}}
    \caption{ Flow field $\mathbf{u}$ around a point torque (rotlet), (a) in an unbounded fluid and (b) in a half-space bounded by a no-slip wall. Cross: location of the rotlet.  Cyan lines: streamlines. Colorbar: $\log|\mathbf{u}|$. The thick line in (b) delimits the region of closed streamlines, while the dashed line indicates the  contour $u_z=0$. $x^*_{12}$ is the separation distance where $u_z=0$ at $z/h=1$.} 
    \label{fig:flow_rotlets}
\end{figure}

Here we propose a dynamical systems approach to study the motion of pairs of microrollers in a viscous fluid, i.e.\ at low Reynolds number.
The dynamics of coplanar neutrally buoyant point microspinners (rotlets) in an unbounded fluid has been  studied theoretically \cite{Leoni2011,Fily2012,lushi2015}. 
A rotlet in an unbounded fluid generates an axisymmetric flow field decaying as $\sim 1/r^2$ (see Fig.\ \ref{fig:flow_rotlets}a). Due to the rotational symmetry of the system, two co-rotating particles with the same torque, $\tau$, always exhibit neutrally stable periodic trajectories regardless of their separation distance $r$. The rotation period  of these trajectories, $T_R$, can be computed analytically and is given by 
\begin{equation}
    T_R = \frac{16\pi\eta r^3}{\tau},
    \label{eq:period_unbounded}
\end{equation}
where $\eta$ is the dynamic viscosity of the suspending fluid, and $\tau$ is the magnitude of the torque applied to the co-rotating particles.
When a no-slip boundary is added to the system, the rotational symmetry is broken. 
The presence of the wall divides the flow around a rotlet into two regions: a recirculating region with closed streamlines around the particle, and another with open streamlines (see Fig.\ \ref{fig:flow_rotlets}b). The size of this recirculating region, in the vertical plane that contains the rotlet center, scales linearly with the height above the floor $h$ (perimeter $\approx 7.1h$, area $\approx 3.7h^2$).  
%\Driscoll{I don't like so many questions in a row so I rewrote this part - this is perhaps more of a style thing, so ignore as you see fit.}  Given the qualitatively different flow field induced by the wall, it is not clear whether or not one can expect to observe periodic states.  We will show that periodic states do exist in the bounded system, and moreover their dynamics is quite rich.  Our calculations include the effects of gravity as well as finite particle size, so that they can easily be mapped to conditions in the lab.  Our results give general conditions under which hydrodynamic bound states may be observed.

Given the qualitatively different flow field induced by the wall, can we then expect to observe periodic states as in the unbounded case ? And, if so, under which conditions ?
 In a more realistic case, how does gravity and  finite particle size affect the dynamics of the system ? 
 More generally, under which conditions does a system, led by hydrodynamic interactions, self-assembles into self-sustained structures ?

A description and the governing equations of the dynamical system is given in Section \ref{sec: Dynamical_sys}. We consider both a simple far-field model using rotlets, as well as a more realistic model that includes gravity, the finite size of the particles and steric repulsion.  We study the equilibrium configurations of these systems and their stability in Section \ref{sec:Phase_space}. In particular we identify the conditions for the emergence of periodic ``leapfrog" trajectories, reminiscent of the particle motion observed in the critters.  Finally, we conclude and discuss how these results may be extended to study the dynamics of a large collection of microrollers at the continuum level in Section \ref{sec:conc_disc}.

\section{Description of the dynamical system}
\label{sec: Dynamical_sys}

\subsection{System and equations of motion}
\begin{figure}
    \centering
    \subfloat[][ ]{\includegraphics[width=0.4\columnwidth]{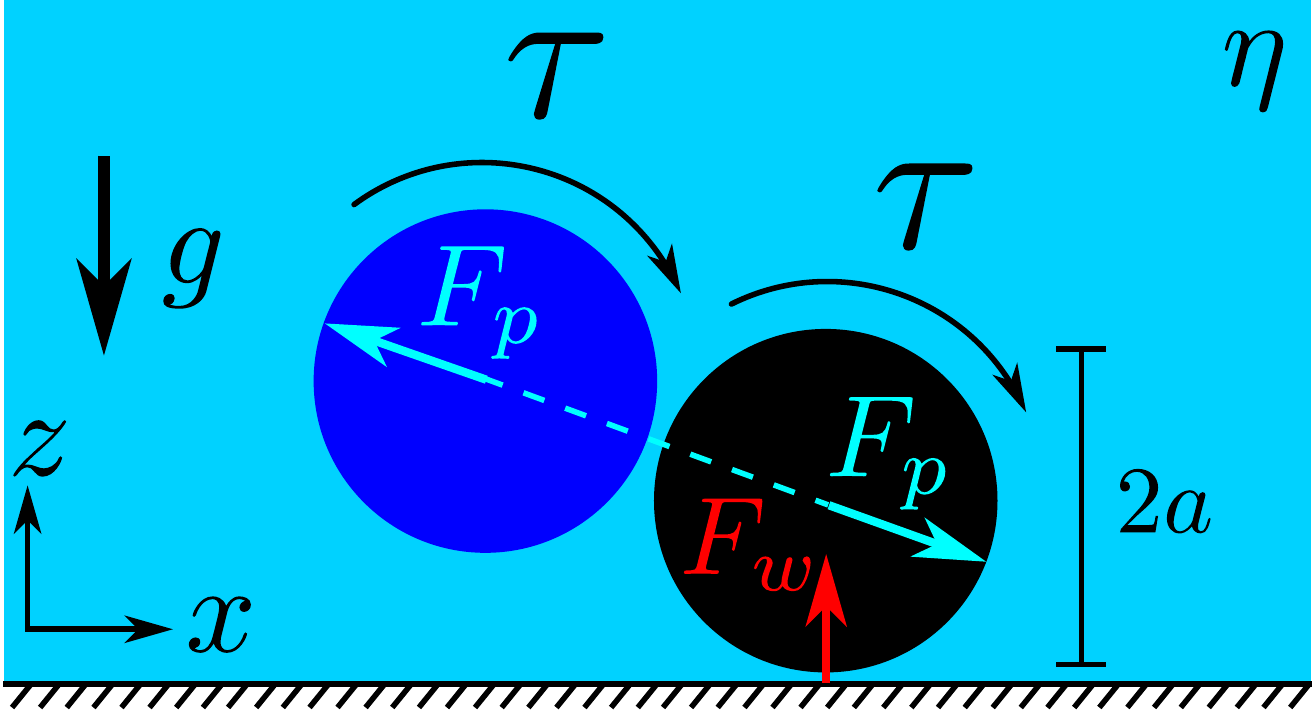}}
    \subfloat[][ ]{\includegraphics[width=0.5\columnwidth]{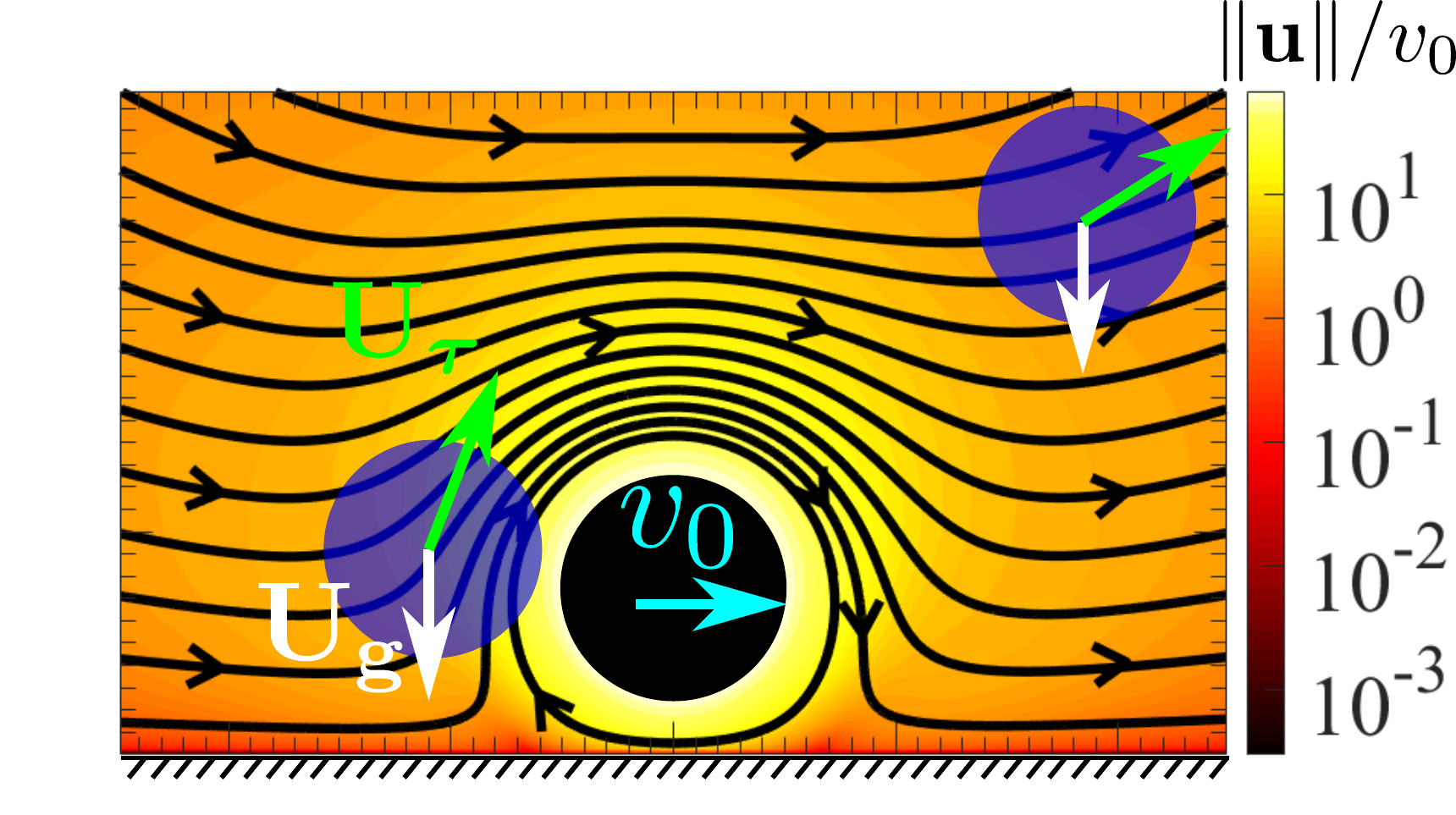}}
    \caption{ (a) Schematic of the system: two particles with radius $a$ rotate under the action of an external torque $\tau$ in a viscous fluid with viscosity $\eta$. They are subject to gravity and inter-particle ($F_p$) and particle-wall ($F_w$) contact forces. (b) Flow field around a torque driven a particle of radius $a$ normalized by the self induced velocity $v_0$, obtained with a high accuracy numerical method \cite{Usabiaga2016,Driscoll2017}. Blue particles represent two different situations where  torque-driven flows can counterbalance or beat gravity to lead to a hydrodynamic bound state. $U_g$ is the self-induced velocity due to gravity and $U_{\tau}$ is the velocity induced by the torque acting on the black particle.} 
    \label{fig:sketch_microrollers}
\end{figure}

We consider two spherical particles, with radius $a$, mass $m$ and coordinates $(x_i,y_i,z_i),\, i=1,2$, where $y_1 = y_2$, above a no-slip boundary located at $z=0$. Each particle is subject to an external torque $\tau=\tau_1=\tau_2$ along the $y$-axis. External forces in the $xz$-plane are written as $2\times 1$ vectors: 
\begin{equation}
    \mathbf{F}_i = \left[\begin{array}{c}
F^x_{p,i}\\
F^z_{p,i} + F_{g,i} + F_{w,i}\\
\end{array}\right],\,\, i=1,2,
\end{equation}
where $\mathbf{F}_p$ is the inter-particle contact force, $F_w$ is the particle-wall contact force and $F_g$ is the gravitational force.
Since the particles are coplanar, their transverse velocity (in the $y$- direction) is zero. See Figure \ref{fig:sketch_microrollers}a for a sketch of the system.

Using the translational invariance of the system along the $x$-axis, one can write the reduced equations of  motion for the state vector $\mathbf{x} = ({x}_{12},{z}_{12},{z}_{C})$

\begin{eqnarray}
\dot{\mathbf{x}} = 
\left[\begin{array}{c}
\dot{x}_{12}\\
\dot{z}_{12}\\
\dot{z}_C
\end{array}\right] &=& \frac{\tau}{\eta  }\boldsymbol{\mu}^{U\tau} + \frac{1}{\eta  }\mathbf{M}^{UF}\cdot \left[\begin{array}{c}
\mathbf{F}_{1}\\
\mathbf{F}_{2}
\end{array}\right] = \mathcal{G}(\mathbf{x}), \label{eq:reduced1}
\end{eqnarray}
where $x_{12} = x_1 - x_2$, $z_{12} = z_1 - z_2$ are the relative positions, and $z_C = (z_1 + z_2)/2$ is the height of the center of mass. $\boldsymbol{\mu}^{U\tau}$ is a $3\times1$ vector of scalar mobility functions that relate the torque applied on the particles along the $y$-axis, $\tau$, to the translational speed of the system $\dot{\mathbf{x}}$. $\mathbf{M}^{UF}$ is the $3\times 4$ mobility matrix that relates the forces applied on the particles $\mathbf{F}_i$ to the translational speed of the system $\dot{\mathbf{x}}$.
$\boldsymbol{\mu}^{U\tau}$ and $\mathbf{M}^{UF}$ contain the self-induced effects as well as hydrodynamic interactions between the particles. These  functions depend exclusively on the  geometric parameters of the system: $\mathbf{x}$ and  $a$.
%$\tilde{x}_{12}=x_{12}/l,\tilde{z}_{12}=z_{12}/l,\tilde{z}_{C}=z_C/l,\tilde{a}=a/l$. 
The RHS of Eq. \ref{eq:reduced1}, $\mathcal{G}(\mathbf{x})$, can be seen as  a vector field that depends on the state vector $\mathbf{x}$.

\subsection{Far-field approximation: rotlets}
In the far field limit, i.e.\ when the particles are far apart from each other and far form the floor, the particles can be modeled as point torques ($a = 0$), called rotlets. Using the rotlet image system above a no-slip boundary from Blake and Chwang \cite{Blake1974}, we derive simple analytic formulas of the mobility functions in terms of the state vector $\mathbf{x}$.
%the caracteristic length can be chosen as the initial height of the center of mass: $l = z_0 = z_C(t=0)$.
In the absence of external forces ($\mathbf{F}_1 = \mathbf{F}_2 = \mathbf{0}$), Eq.\ \ref{eq:reduced1} reduces to a simple expression
\begin{eqnarray}
\dot{\mathbf{x}} = 
\left[\begin{array}{c}
\dot{x}_{12}\\
\dot{z}_{12}\\
\dot{z}_C
\end{array}\right] &=& 
\left[\begin{array}{c} 2z_{12}\left(\frac{1}{r_{12}^3} - \frac{1}{R_{12}^3} +  3\frac{x_{12}^2}{R_{12}^5}\right) \\
 2x_{12}\left(\frac{1}{R_{12}^3} - \frac{1}{r_{12}^3} +  12\frac{z_C^2}{R_{12}^5}\right) \\
 6\frac{x_{12}z_{12}z_C}{R_{12}^5}.
 \end{array}\right] = \mathcal{G}^{\mbox{rotlet}}(\mathbf{x}),
\label{eq:rotlets_reduced}
\end{eqnarray}
where  $r^2_{12} = x_{12}^2 + z_{12}^2$ and $R^2_{12} = x_{12}^2 + 4z_C^2$. Here length and time have been nondimensionalized by the initial height of the system $l_c = z^0_C = z_C(t=0)$ and $t_c = \eta l_c^3/\tau$ respectively.
Since the rotlets cannot cross the floor, the admissible phase space is bounded by the planes $z_{12} = \pm 2z_C$. As shown by Eq. \eqref{eq:rotlets_reduced}, this constraint is automatically satisfied since $\dot{z}_{12} = \pm 2\dot{z}_C$ when $z_{12} = \pm 2z_C$.

%Note that contrary to coplanar rotlets in an ubounded fluid \cite{lushi2015}, or point vortices \cite{Aref1983}, the system \eqref{eq:rotlets_reduced} does not have a Hamiltonian structure. Therefore we cannot find a \Driscoll{Say what this implies/why you are pointing this out}

%\begin{eqnarray}
%\dot{x}_1 &=& z_{12}\left(\frac{1}{r_{12}^3} - \frac{1}{R_{12}^3}\right) +  6\frac{x_{12}^2z_1}{R_{12}^5}\\
%\dot{x}_2 &=& -z_{12}\left(\frac{1}{r_{12}^3} - \frac{1}{R_{12}^3}\right) +  6\frac{x_{12}^2z_2}{R_{12}^5}\\
%\dot{z}_1 &=& x_{12}\left(\frac{1}{R_{12}^3} - \frac{1}{r_{12}^3}\right) +  6\frac{x_{12}z_1(z_1+z_2)}{R_{12}^5}\\
%\dot{z}_2 &=& -x_{12}\left(\frac{1}{R_{12}^3} - \frac{1}{r_{12}^3}  \right) -6\frac{x_{12}z_2(z_1+z_2)}{R_{12}^5},
%\end{eqnarray}

\subsection{Particles with finite radius}
\label{sec:Finite-sized particles}
When  the inter-particle or  particle-wall distance is of the order of the particle radius, $a$, the finite-size effects must be included. %\Driscoll{Why did we not have to worry abou thtis in the other papers - was it because we used a continuum model and assumed a density?} 
We account for these effects by using the Rotne-Prager-Yamakawa mobility \cite{Rotne1969,Yamakawa1970} with wall corrections derived by Swan and Brady (see Appendices B and C in \cite{Swan2007}). To obtain an equation in the form of Eq. \ref{eq:reduced1}, we rearrange the formulas from \cite{Rotne1969,Yamakawa1970,Swan2007} in terms of $\mathbf{x} = ({x}_{12},{z}_{12},{z}_{C})$ and $a$. %\Driscoll{I would put these in an appendix}
 As shown in Figure \ref{fig:sketch_microrollers}b, spheres of radius $a$ rotating above a floor translate with a self induced horizontal velocity $v_0$, which, to leading order, scales as $(a/h)^4$ \cite{Swan2007}, while rotlets do not have a self-induced velocity.

After nondimensionalizing Eq. \ref{eq:reduced1} with $l_c = a$ \footnote{we could have chosen $l_c=z^0_C$ but, as shown below, the phase space is simpler to visualize with $l_c=a$.} and $t_c = \eta l_c^3/\tau$, we obtain 
%If the particles are not touching each other nor the wall, i.e. when contact forces vanish, then $\mathbf{F}_1 = \mathbf{F}_2 = [0,  -mg]^T$, and the resulting nondimensionalized equations are given by
\begin{eqnarray} 
\dot{\mathbf{x}} = 
\left[\begin{array}{c}
\dot{x}_{12}\\
\dot{z}_{12}\\
\dot{z}_C
\end{array}\right] &=& \boldsymbol{\mu}^{U\tau} + B^{-1} \mathbf{M}^{UF}\cdot 
\left[\begin{array}{c}
\mathbf{F}_{1}\\
\mathbf{F}_{2}
\end{array}\right], \label{eq:reduced_B}
\end{eqnarray}

where $B = \tau/(mga)$ is a dimensionless number that compares the strength of gravity  with external torques.  In the limit $B\rightarrow 0$, the dynamics is mainly dictated by gravity, while when $B\rightarrow \infty$ the system is dominated by  the torque-driven flows. 
 $B$ can be seen as the ratio of the self-induced velocity due to gravity and the velocity induced by the torque acting on another particle (see Fig. \ref{fig:sketch_microrollers}b): $B = U_{\tau}/U_g$ where $U_{\tau} \sim \tau/\eta a^2$ and $U_g \sim mg/\eta a$. When $B\gg1$, the upward torque-induced flows can counterbalance or overcome gravity and lead to  hydrodynamic bound states. If one assumes a critter-like structure with typical size $L$,  $B$  can also be defined as the ratio of two characteristic times: $B = t_g/t_\tau$ where $t_g = L/U_g \sim L\eta a/mg$ is the characteristic time for a particle to fall across and exit the critter, and $t_\tau \sim L/U_\tau =  L\eta a^2/\tau$ is the ``overturn" time, i.e.\ the time for a particle to travel the perimeter of the critter. So a particle is more likely to stay in the critter when $t_\tau \ll t_g$, that is when $B\gg 1$.

In this work, we want to focus only on the effect of $B$ on the system. 
Here, the goal of repulsive contact forces is to prevent overlaps while having as little influence as possible on the dynamics of the system.  We therefore chose to use extremely short-range repulsive forces; in practice, we model contact forces with an exponentially decaying repulsive potential of the form \cite{Usabiaga2016b}
\begin{equation}
    U(r) = \begin{cases}
U_0 (1+\frac{d-r}{b}) & \mbox{if } r<d,\\
U_0\exp(\frac{d-r}{b})  & \mbox{if } r\ge d.\\
\end{cases}
\label{eq:potential}
\end{equation}
For particle-particle interactions, $r$ is the center-to-center distance and $d=2a$. For particle-wall interactions, $r$ is the height of the particle center and $d=a$. The energy scale $U_0$ and interaction range $b$ control the strength and decay of the potential respectively. 
We found that taking $U_0 = \max(3mga/2,\tau/20)$ and $b=0.025a$ prevent particle-particle and particle-wall overlaps while keeping close contact. 

Since spheres cannot overlap each other nor the wall, the admissible phase space is bounded by the planes $z_C=a$, $z_{12} = \pm 2(z_C-a)$ and the cylinder $x^2_{12}+z^2_{12} = 4a^2$. %\Driscoll{I find the Figure showing this not very helpful and I suggest eliminating it}. 
 This region is  discretized using a polar cylindrical mesh. The vector field $\mathcal{G}(\mathbf{x})$ (Eq. \ref{eq:reduced1}) is  evaluated at the mesh nodes and, streamlines, i.e.\ trajectories, are analyzed.

% \begin{figure}
%     \centering
%     \includegraphics[width=0.65\columnwidth]{Admissible_phase_space_with_mesh.png}
%     \caption{Admissible phase space for two finite particles. The bounding  surface  $S=\left\{\mathbf{x}=(x_{12},z_{12}, z_C) \,| \,x^2_{12}+z^2_{12} = 4a^2, z_C = a, z_{12} = \pm 2(z_C-a)\right\}$ is colored in light grey. The blue lines represent a slab of the mesh in cylindrical coordinates (the actual mesh is much finer).} 
%     \label{fig:admissible_phase_space}
% \end{figure}

\section{Dynamics in phase space}
\label{sec:Phase_space}

In this section we analyze, numerically and analytically, the trajectories and their stability in phase space.  First, we consider the limit $B\rightarrow \infty$, where the system is only driven by active flows. Then, we study the evolution of the attractors and limit cycles of the system as $B$ varies. We show that the system undergoes various bifurcations and that periodic orbits can be obtained above a threshold value $B>B^*$.

\subsection{Neutrally stable states: $B \rightarrow \infty$}
When $B\rightarrow \infty$ the particles can be considered as neutrally buoyant, thus the dynamics of the system is only driven by the active flows induced by the external torque.  Below we analyze the resulting neutrally stable states and compare the rotlet solutions with the finite sphere solutions.

\subsubsection{Rotlets}
\begin{figure}
    \centering
    \includegraphics[width=0.8\columnwidth]{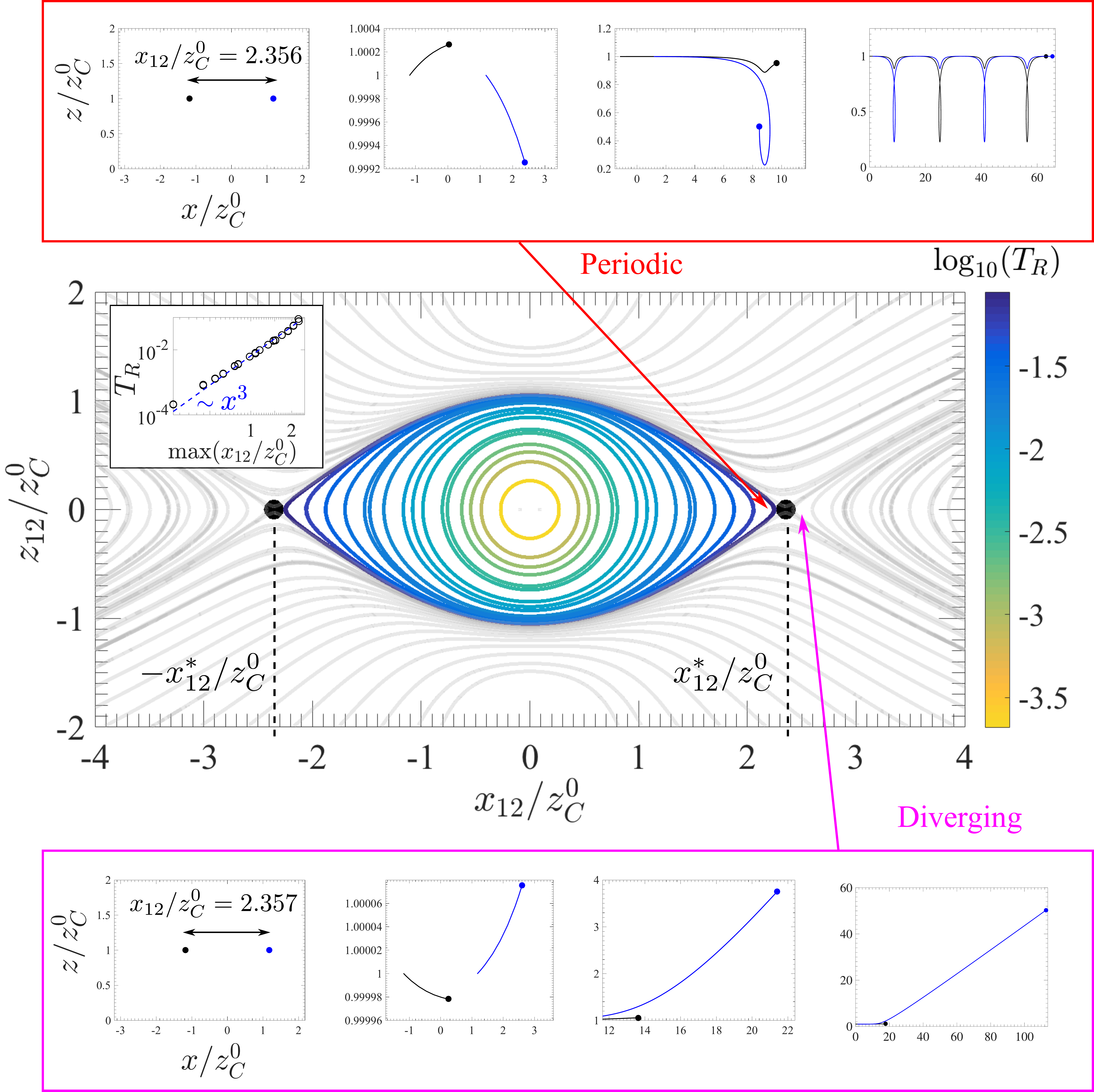}
    \caption{Trajectories of two rotlets in phase space for $B=\infty$ colored by their period $T_R$. Grey trajectories  diverge. The black circles represent the points $(x_{12}^*/z_C^0 = \pm 2.356927998,\, z_{12}/z_C^0 = 0, \, z_{C}/z_C^0 = 1)$ that lie on the separatrix between neutrally stable and unstable states. Inset: period of trajectories as a function of the maximum horizontal separation, the dashed blue line is a fit: $T_R \sim \max(x_{12}/z_C^0)^3$. Upper and lower panels: trajectories in physical space chosen in the vicinity of $x_{12}^*$ ($x_{12}^0/z_C^0 = 2.356$ and $x_{12}^0/z_C^0 = 2.357$ respectively).  } 
    \label{fig:rotlet_phase_space}
\end{figure}

The trajectories of the autonomous dynamical system \eqref{eq:rotlets_reduced} are highly sensitive to initial conditions.
The upper and lower panels in Figure \ref{fig:rotlet_phase_space} show an example for two rotlets initially placed  at the same height: $z_{12}^0/z_C^0 = 0$. When their initial separation distance is 
  $x_{12}^0/z_C^0 = 2.356$, the rollers exhibit a periodic leapfrog motion, while a small increment,  $x_{12}^0/z_C^0 = 2.356 + 0.001 = 2.357$, leads to a diverging trajectory.
  The critical separation value $x_{12}^*$ for two particles 
  at the same height is given by $\dot{z}_{12}=0 \implies 2x_{12}\left(\frac{1}{R_{12}^3} - \frac{1}{r_{12}^3} +  12\frac{z_C^2}{R_{12}^5}\right) = 0 \implies x_{12}^*/z_C^0 = \pm 2.356927998$ (see Figure \ref{fig:flow_rotlets}b) \footnote{Note that $\dot{x}_{12}=0$ and $\dot{z}_{C}=0$ when $z_{12}=0$. It is also worth mentioning that $x_{12} = 0$ is a singular critical point.}. As shown in the lower and upper panels of Figure \ref{fig:rotlet_phase_space}, when $x_{12}>x_{12}^*$, the rear particle heads downward and the front particle is lifted upward. Such situation is unstable and leads to a diverging trajectory. When $x_{12}<x_{12}^*$, the rear particle goes up and the front particle goes down, which then leads to a periodic leapfrog motion.

Figure \ref{fig:rotlet_phase_space} shows the trajectories in phase space ($x_{12},z_{12},z_{C}$), rescaled by the initial height $z_C^0$, for various initial separations ($x_{12}^0/z_C^0$ and $z_{12}^0/z_C^0$), colored by period of rotation. 
Contrary to the unbounded case (Eq. \ref{eq:period_unbounded}), the period of rotation no longer scales with $r^3$ because to the loss of symmetry in the system, but, as shown in the inset, seems to follow a cubic scaling with the maximum horizontal separation. The phase space exhibits a discontinuous transition between the region of neutrally stable states and unstable trajectories. 
%\Driscoll{How do you know the transition is discontinuous?}. 
This discontinuity is purely hydrodynamic in origin and is related to the geometry of the flow field (Figure \ref{fig:flow_rotlets}b): any trajectory that passes through a point with coordinates $(|x_{12}|/z_C^0<x^*_{12},0,z_C/z_C^0 = 1)$ is neutrally stable and periodic, in other words, every time the particles reach the same height their separation distance must be less than $x_{12}^*$ to remain in a periodic hydrodynamic bound state. If the particles never reach the same height, then the trajectory diverges. %\Driscoll{Are you making a statement about whether the torque-generated  flow is enough to keep the pair in the bounded parts of their respective flows - be more specific here.} 
Note that this system exhibits a phase space similar to the well-known pendulum without friction \cite{Stokes1851,Baker2005}, except that the neutrally stable trajectories do not replicate periodically in phase space.

\subsubsection{Torque-driven spheres}
\begin{figure}
    \centering
    %\subfloat[][Trajectories colored by length of periods.]{\includegraphics[width=0.5\columnwidth]{velocity_paraview_polar_fitted_zcmin_1_0001_6a_normalize_h_0_Nt_Nr_Nzc_101_100_150_B_Inf_streamlines_periods_3D.png}}
    \includegraphics[width=0.8\columnwidth]{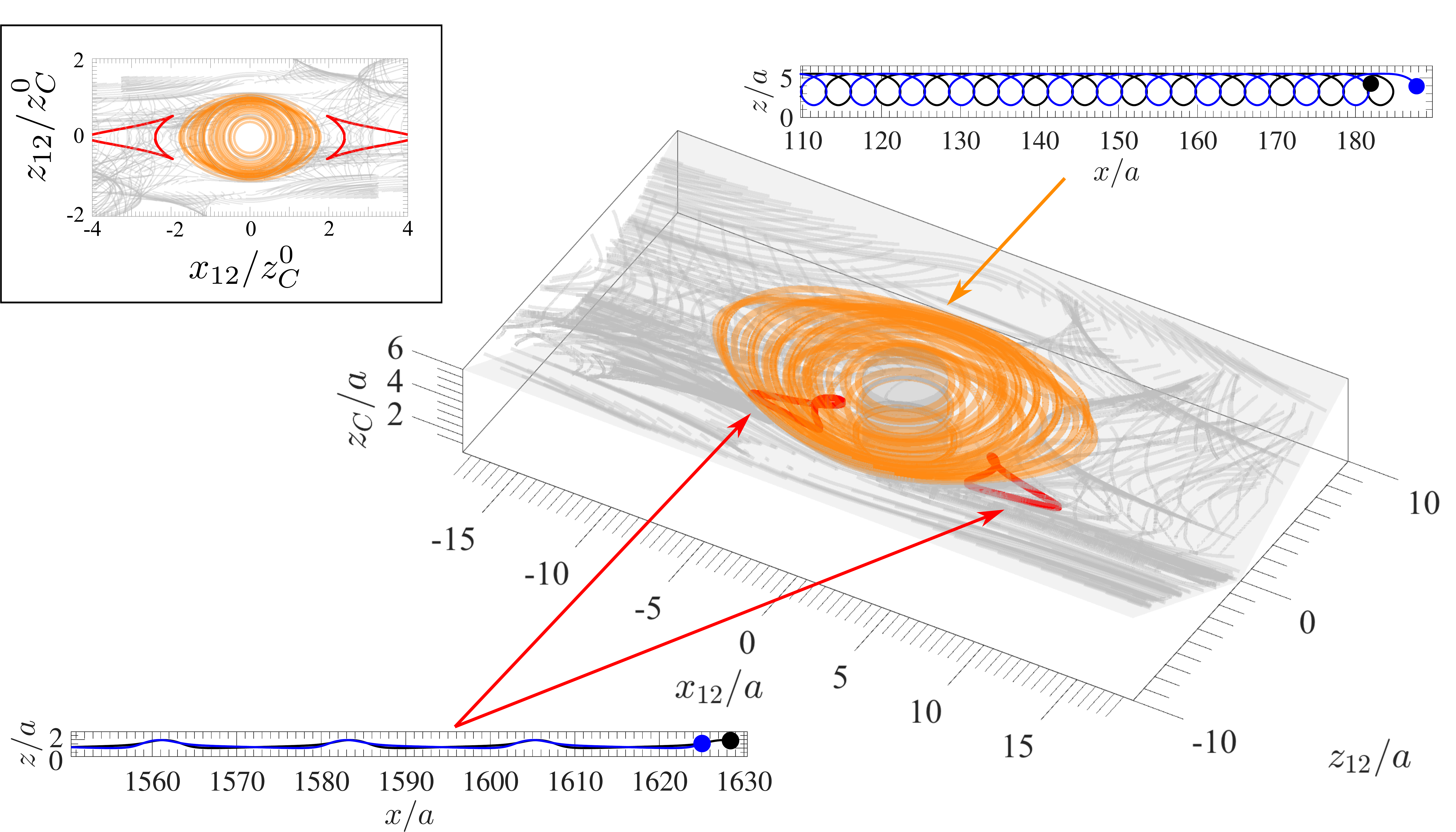}
    \caption{Trajectories of two spheres of radius $a$ in phase space rescaled by  $a$ for $B=\infty$. Trajectories colored by type of periodic motion. Orange trajectories: leapfrog motion; red trajectories: vertical periodic  oscillations. Grey trajectories diverge. Inset: trajectories rescaled by the initial height of the system $z_C^0$. } %\Driscoll{I would rotate these more side-on, it is really hard to see what you are pointing out.  Additionally, do you need panel a? - you don't say anything about it.} 
    \label{fig:Binf_phase_space}
\end{figure}

Similarly to a rotlet, a torque-driven sphere, with radius $a$, above a floor generates a recirculating region whose size depends on $h/a$  (cf. Figure 1c in \cite{Driscoll2017}).
Figure \ref{fig:Binf_phase_space}  shows  the trajectories of spheres colored by type and rescaled by $a$. As shown in the inset, at large distances, the behavior of the system is similar to the pair of rotlets: when rescaling lengths by $z_C^0$ instead of $a$, the trajectories collapse  as in Figure \ref{fig:rotlet_phase_space}.  However, finite size corrections are perceptible closer to the wall.  In addition to the leapfrogging trajectory (orange), a qualitatively different periodic trajectory appears (red) for $z_C<2a$.   The new periodic trajectory corresponds to a vertical oscillation of the translating pair of microrollers. Without the self-induced velocity due to the finite size of the spheres, this trajectory would diverge.

%\begin{figure}
%    \centering
   
    %\subfloat[][]{\includegraphics[width=0.95\columnwidth]{velocity_rotlet_paraview_polar_fitted_zcmin_1_0001_6a_Nt_Nr_Nzc_101_100_150_B_Inf_streamlines_periods_3D.png}}
%    \caption{ } 
%    \label{fig:Binf_phase_space2}
%\end{figure}

\subsection{A rich phase space for a simple system: $0<B<\infty$}

As seen in the previous section, when $B\rightarrow \infty$, all the periodic trajectories are neutrally stable; there is no attractor in the system.
In the opposite regime, $B=0$, the system is exclusively driven by gravity and there are infinitely many sinks  with coordinate $z_C = z_C^{\mbox{eq}}$, where $z_C^{\mbox{eq}}$ is the equilibrium value of $z_C$ obtained by balancing  gravity and contact forces with the floor. $z_C^{\mbox{eq}}=a$ for a hard-sphere potential and $z_C^{\mbox{eq}} \approx 1.1a$ for our repulsive potential \eqref{eq:potential}.

\begin{figure}
    \centering
    \includegraphics[width=0.95\columnwidth]{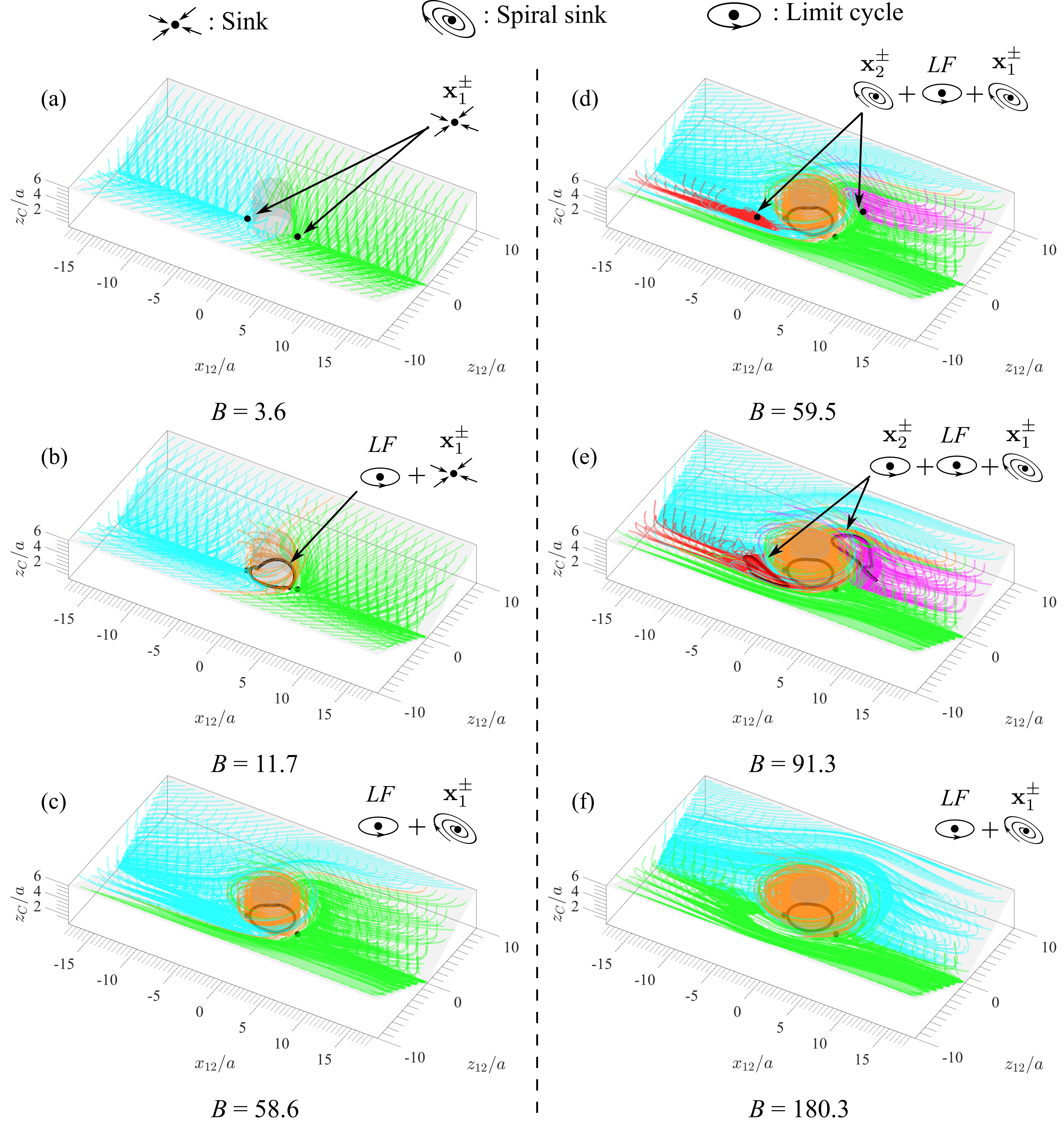}    \caption{Trajectories of two  particles in phase space for finite  values of $B$. Black circle: fixed points (sinks). Black trajectories: limit cycles. Each color corresponds to different basin of attraction: blue and green trajectories are attracted to the fixed points $\mathbf{x}_1^{\pm}$; orange trajectories converge to the leapfrog  limit cycle ($LF$); red and magenta trajectories are attracted to the fixed points $\mathbf{x}_2^{\pm}$, that undergoes a Hopf bifurcation at $B=77$.} 
    \label{fig:B_phase_space}
\end{figure}

Between these two limits, the landscape  of the phase space evolves between six regimes:
\begin{enumerate}
    \item $B<11.1$: two sinks (Fig. \ref{fig:B_phase_space}a)
    \item $11.1<B<12.8$: one limit cycle + two sinks (Fig.\ \ref{fig:B_phase_space}b)
    \item $12.8<B<59.5$: one limit cycle + two spiral sinks (Fig.\ \ref{fig:B_phase_space}c)
    \item $59.5<B<77$: one limit cycle + four spiral sinks (Fig.\ \ref{fig:B_phase_space}d)
    \item $77 < B \le 95.5$: three limit cycles + two spiral sinks (Fig.\ \ref{fig:B_phase_space}e)
    \item $95.5<B<\infty$: one limit cycle + two spiral sinks (Fig.\ \ref{fig:B_phase_space}f)
\end{enumerate}

Figure \ref{fig:B_phase_space} shows the phase space for the six regimes. Trajectories are colored according to the basin of attraction to which they belong.

Below we identify the fixed points and limit cycles of these regimes and analyze their evolution as $B$ varies.

\subsubsection{Leapfrog orbits (LF)}
\begin{figure}
    \centering
    \includegraphics[width=0.55\columnwidth]{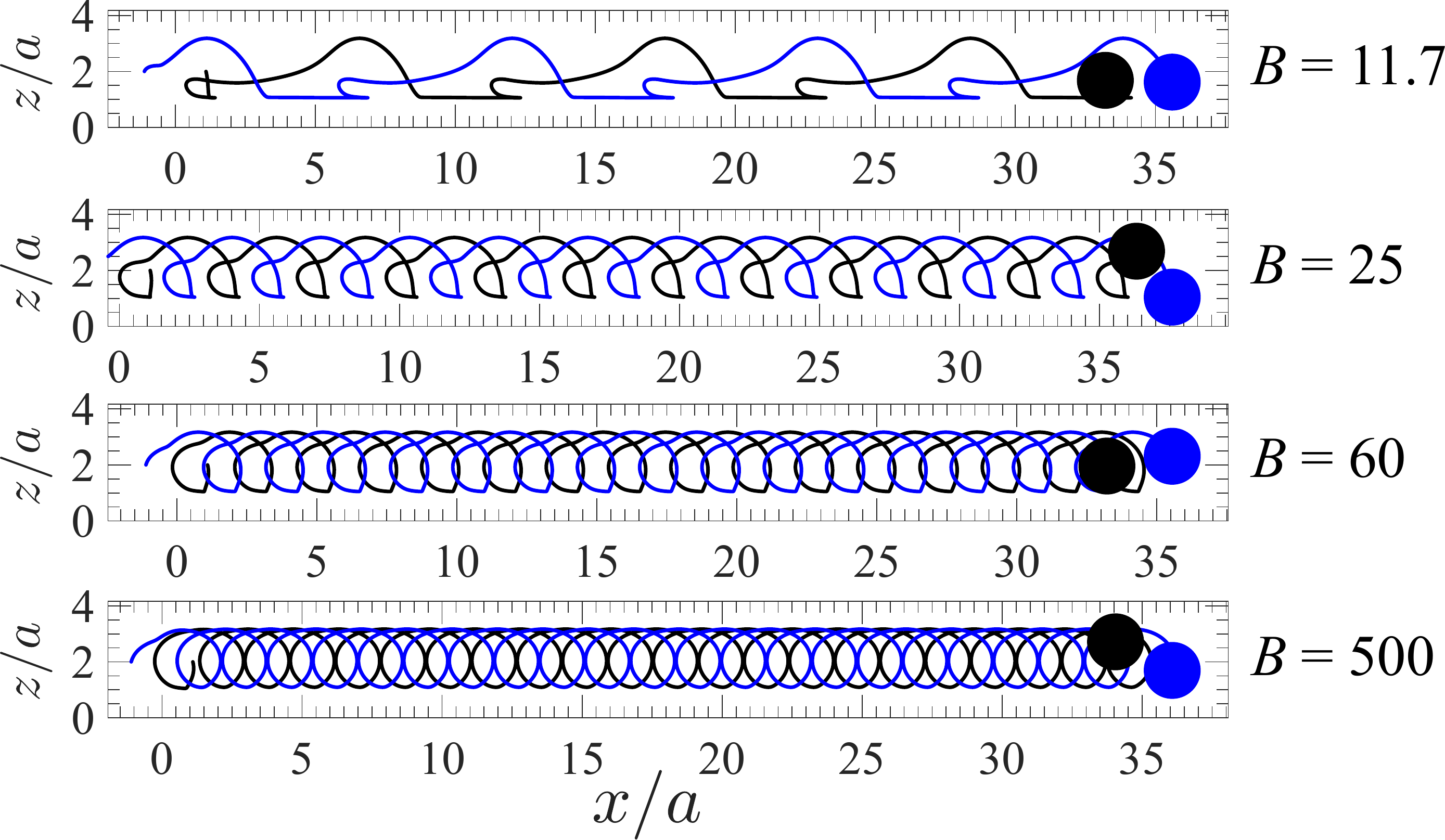}    \caption{Leapfrog trajectories for $B=11.7-500$, where $B^*=11.7$ is the critical value. } 
    \label{fig:traj_leapfrog}
\end{figure}

Unlike neutrally buoyant particles, the existence of leapfrog trajectories depends not only on the initial particle positions but also on the competition between gravity and torque-driven flows. 
This competition defines a critical value, $B^*$, above which active flows overcome the effect of gravity and a leapfrog trajectory appears (see Fig. \ref{fig:sketch_microrollers}b). This discontinuous transition is shown  on Figure \ref{fig:B_phase_space}b as the appearance of the orange trajectories that converge to the leapfrog limit cycle (thick black line).  The leapfrog trajectory is a unique limit cycle that appears at $B^*=11.7$, and whose basin of attraction increases with $B$. 
Figure \ref{fig:traj_leapfrog} shows the evolution of the leapfrog trajectories as $B$ increases. When $B=B^*=11.7$, active flows are just strong enough to overcome the delaying effect of gravity (cf.\ Fig.\ \ref{fig:sketch_microrollers}b), as shown by the shape of the trajectories.  As $B$ increases, the period of the orbit decreases and the motion becomes circular.

The leapfrog orbit is reminiscent of the periodic treadmilling motion followed by the microrollers that self-assemble into critter (Fig. \ref{fig:critters}).  Therefore, for a given particle size and mass, these results indicate that critters can only appear above a threshold value of the external torque, which is also suggested by our previous experimental observations and numerical simulations \cite{Usabiaga2016b,Driscoll2017}.
 %\Driscoll{Can you compute a value for $B$ that is at all in agreement with the experimental/numerical results?  I remember you had some slides on this a while back, but I am not sure where that went.}
Various parameters can be changed in the experiments to vary $B$, for example particle size, magnitude of the external torque, and particle mass.  We will explore these directions in the near future to better understand how the critter state emerges, and how it can be controlled.
%to optimize  critters for microfluidic transport.

%These critters are observed when the particles are high enough to generate a large recirculating motion. The torque-driven flows induced by the microllers lift particles off the floor  to form a layer \cite{Usabiaga2016b} whose height is set by $B$: the larger $B$, the higher the particles, and thus the more likely they are to form critters.

\subsubsection{Surviving sinks $\mathbf{x}_1^{\pm}$}
\begin{figure}
    \centering
    \subfloat[][Eigenvalues $\lambda_{1,2,3}$ of $\nabla \mathcal{G}(\mathbf{x}_1^{\pm})$ \textit{vs.} $B$ (cf. Eq. \ref{eq:linearized}). Each color represents a different eigenvalue. ]{\includegraphics[width=0.4\columnwidth]{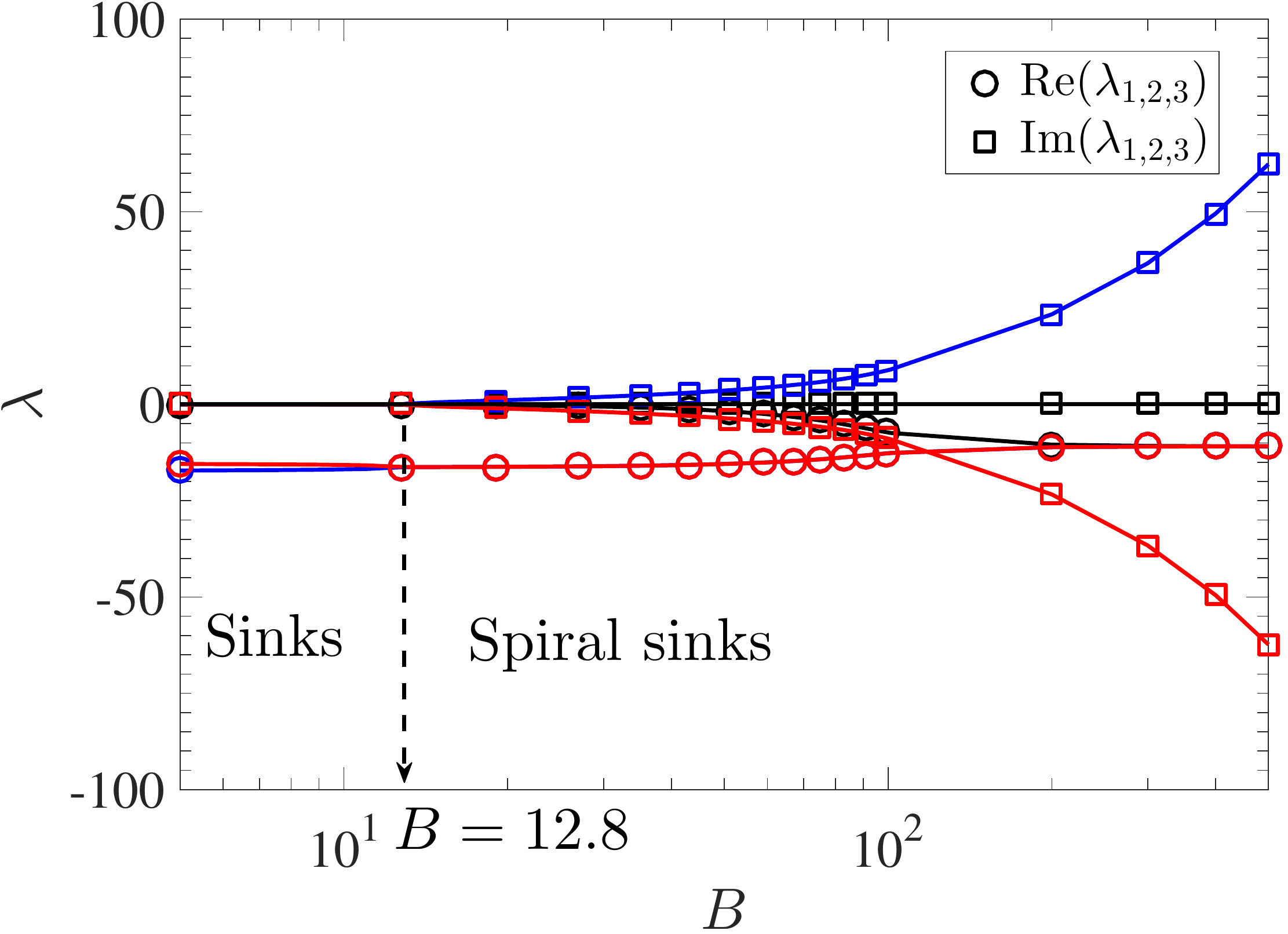}}
    \hspace{0.5cm}
    \subfloat[][Linearized trajectories in the vicinity of $\mathbf{x}_1^{\pm}$ colored by the value of $B$. Black circle: position of  $\mathbf{x}_1^{\pm}$.]{\includegraphics[width=0.4\columnwidth]{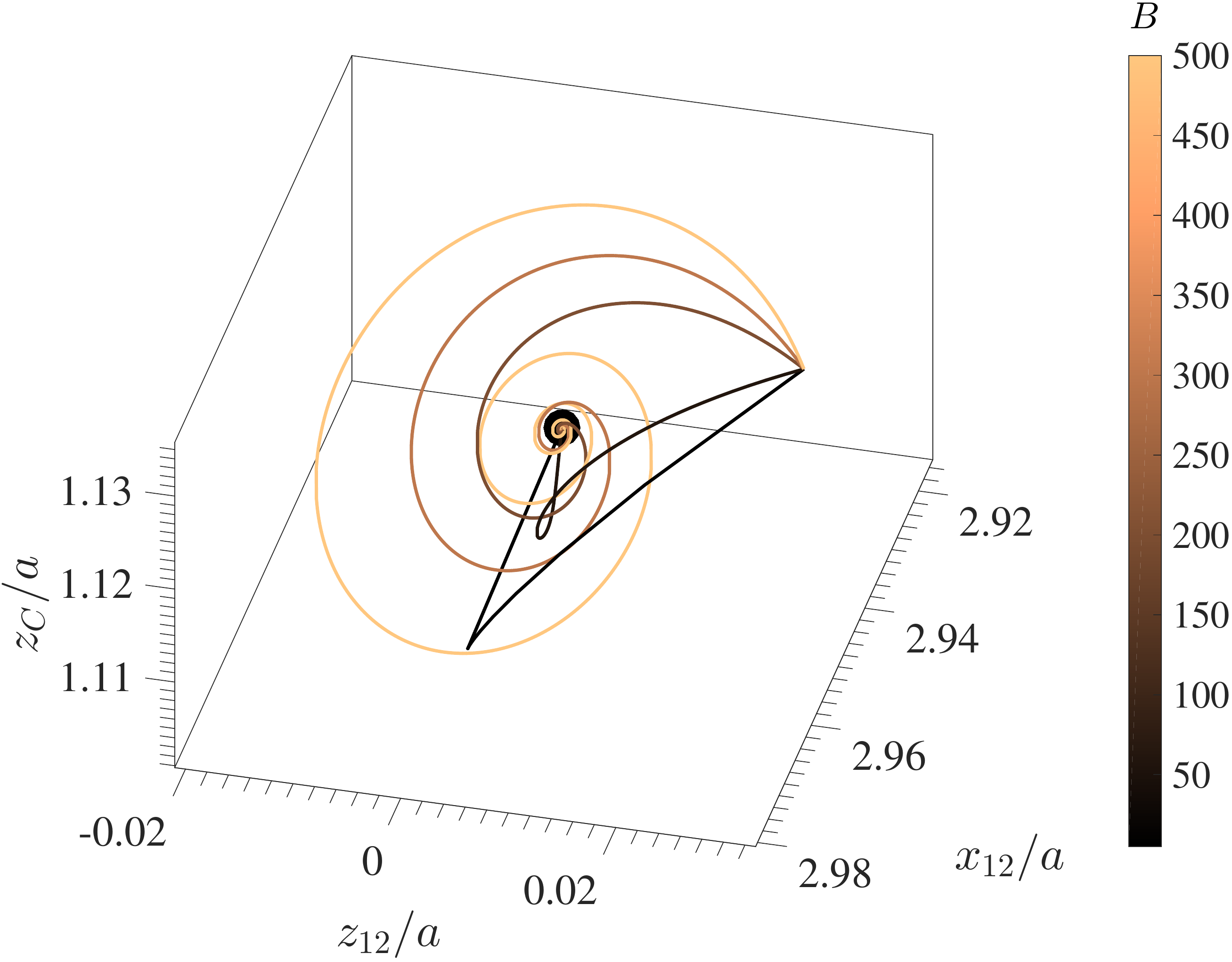}}\\
    \subfloat[][Examples of trajectories converging to $\mathbf{x}_1^{\pm}$ in physical space.]{\includegraphics[width=0.5\columnwidth]{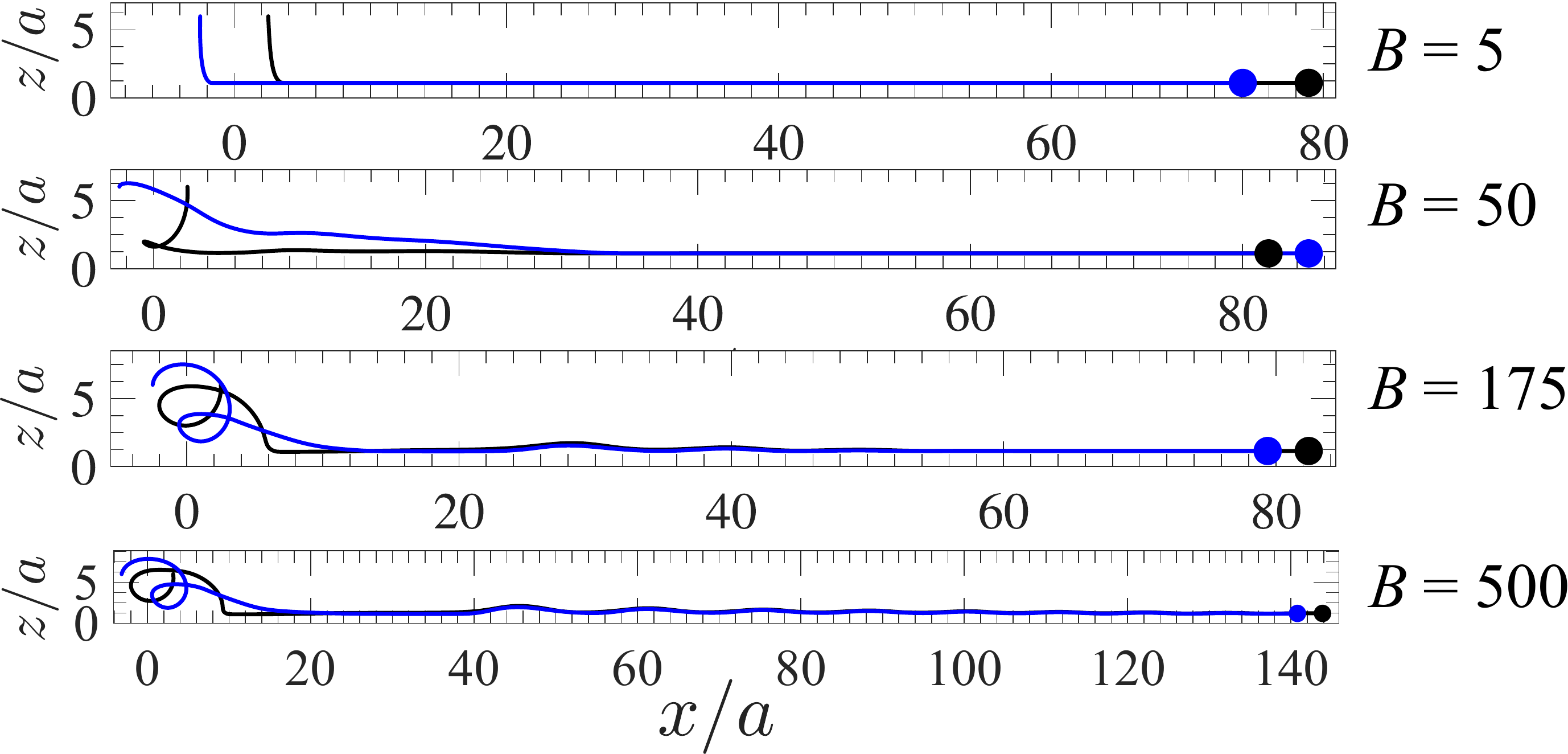}}
    \caption{Characterization of the surviving sinks $\mathbf{x}^{\pm}_1$ } 
    \label{fig:B_phase_space_fixed_point_1}
\end{figure}

Two symmetric sinks  survive for all values of $B \in  [0;\infty[$, these are indicated in Fig.\ \ref{fig:B_phase_space}a-f by the black circles close to the cylinder delimiting the excluded volume region. As shown in Figure \ref{fig:B_phase_space_fixed_point_1}c, these fixed points, with coordinates $\mathbf{x}_1^{\pm} = (\pm 2.95a,0,z_C^{\mbox{eq}})$, correspond to a stable state where the particles are sedimented near the floor  and translate at the same height  with a constant separation distance ($x_{12}= \pm 2.95a$ ).  To study the stability of these fixed points, we linearize
Eq. \ref{eq:reduced1}  about  $\mathbf{x}_1^{\pm}$ 
\begin{eqnarray}
 \delta\dot{\mathbf{x}}_1 &=& \nabla \mathcal{G}(\mathbf{x}_1^{\pm})\cdot\delta\mathbf{x}_1.
 \label{eq:linearized}
\end{eqnarray}
where $\delta\mathbf{x}_1 = \mathbf{x} - \mathbf{x}_1^{\pm}$.  Note that $\nabla \mathcal{G}(\mathbf{x}_1^{\pm})$ is expressed \textit{analytically} as a function of $B$ and the contact force parameters. As mentioned in Section \ref{sec:Finite-sized particles}, the contact force parameters are chosen to minimize their influence on the trajectories of the system.
The three eigenvalues and eigenvectors of $\nabla \mathcal{G}(\mathbf{x}_1^{\pm})$
are then computed.

Figure \ref{fig:B_phase_space_fixed_point_1}a shows the real and imaginary part of each eigenvalues for $B=1-500$. First, we note that all eigenvalues have negative real parts, confirming that this is a stable fixed point. We also observe a bifurcation  from zero to  non-zero (conjugate) imaginary parts  at $B=12.8$, where $\mathbf{x}_1^{\pm}$ becomes a spiral sink.   Figure \ref{fig:B_phase_space_fixed_point_1}b shows the corresponding linearized trajectories in phase space in the vicinity of $\mathbf{x}_1^{\pm}$ for $B=1-500$. Below the threshold value $B=12.8$, the system relaxes exponentially to the fixed point. As $B$ increases, the active flows get stronger and the trajectories oscillate towards  $\mathbf{x}_1^{\pm}$ (see third  and fourth panel in Fig. \ref{fig:B_phase_space_fixed_point_1}c).  Note that $x_{12}$ is the slowest degree of freedom, which is intuitive since gravity tends to stabilize $z_{12}$ and $z_C$ first. Once $z_{12}$ and $z_C$ are close to their equilibrium value, the only non-zero component of the vector field is the first component,  $\mathcal{G}_x$, along the $x_{12}$-axis. $\mathcal{G}_x(x_{12},0,z_C^{\mbox{eq}})$ is very small and has only one zero at $x_{12} = \pm C$. Therefore, when a pair of particle settles to the floor, i.e.\ when $z_{12}\rightarrow 0$ and $z_C \rightarrow z_C^{\mbox{eq}}$, it moves slowly in phase space along the $x_{12}$-axis towards  $\mathbf{x}_1^{\pm}$ (see Fig. \ref{fig:B_phase_space_fixed_point_1}b-c) . 
Thus, compared to the other fixed points, $\mathbf{x}_1^{\pm}$ has the largest basin of attraction

\subsubsection{Hopf bifurcation at $\mathbf{x}_2^{\pm}$}
\begin{figure}
    \centering
     \subfloat[][Eigenvalues $\lambda_{1,2,3}$ of $\nabla \mathcal{G}(\mathbf{x}_2^{\pm})$ \textit{vs.} $B$. Black lines: conjugate complex eigenvalues $\lambda_{1,2}$. Blue line: real eigenvalue $\lambda_{3}$.]{\includegraphics[width=0.4\columnwidth]{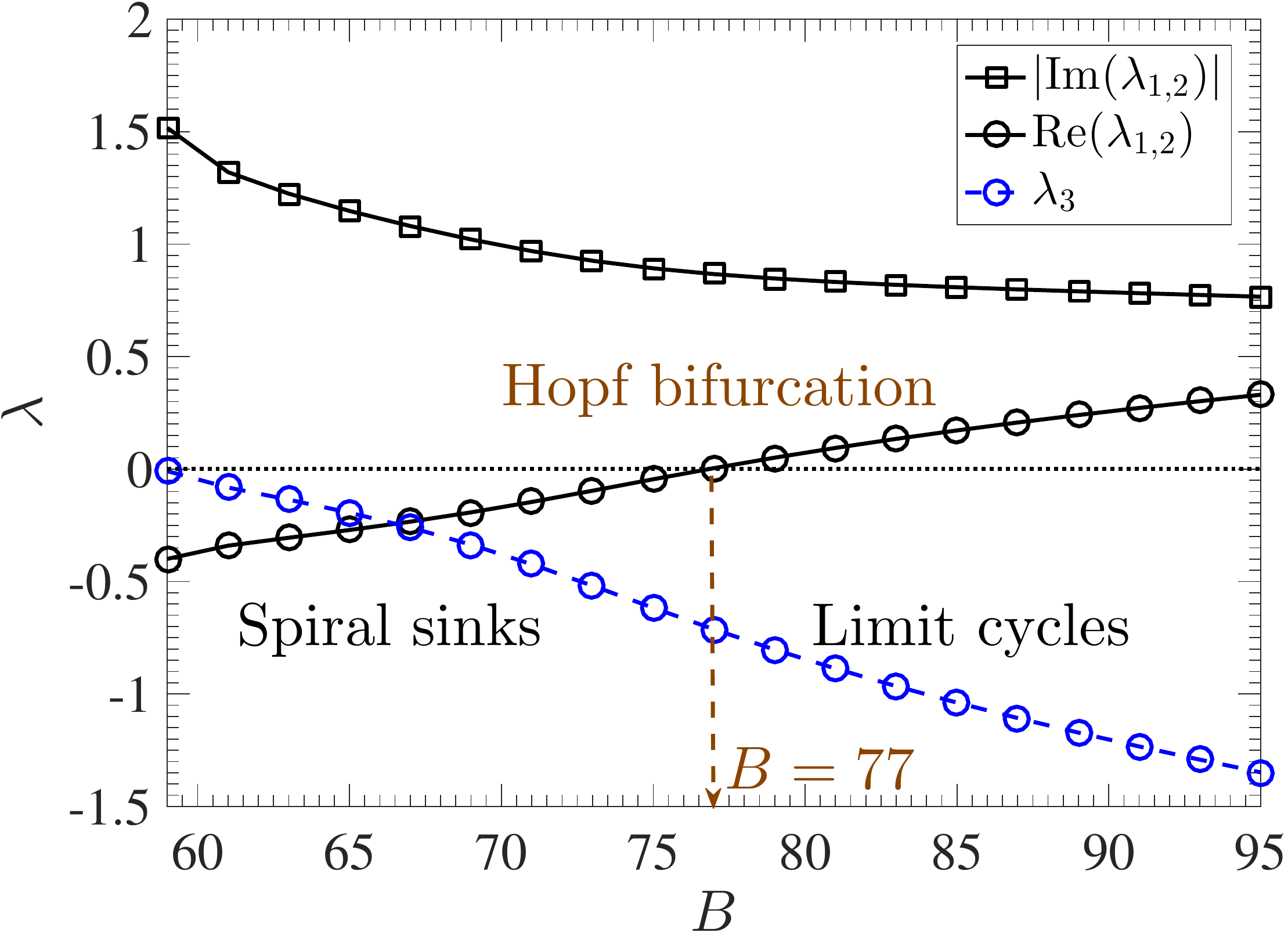}}
     \hspace{0.5cm}
    \subfloat[][Linearized trajectories in the vicinity of $\mathbf{x}_2^{\pm}$ colored by the value of $B$. Circles: position of  $\mathbf{x}_2^{\pm}$.]{\includegraphics[width=0.4\columnwidth]{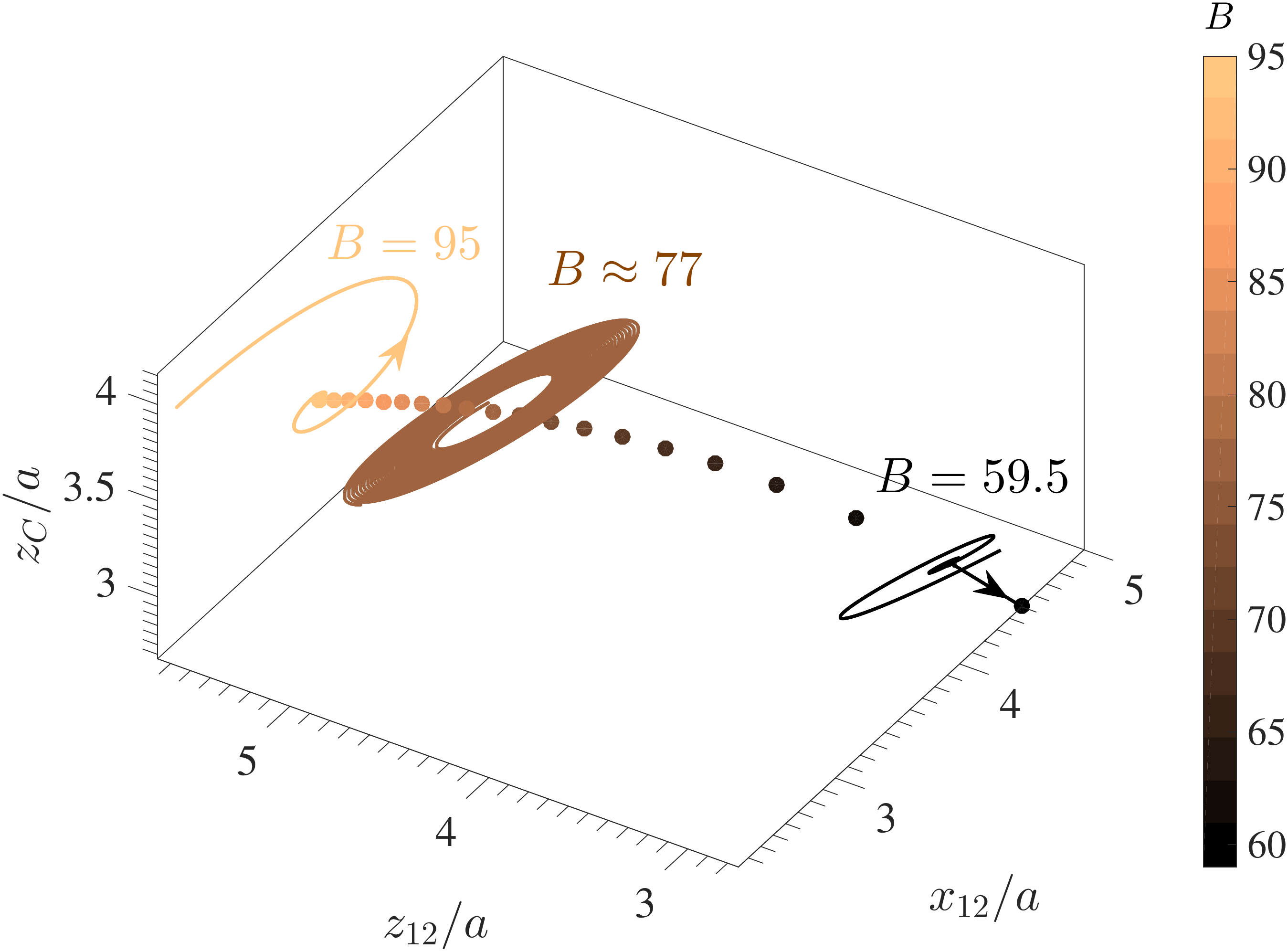}}\\
     \subfloat[][Examples of trajectories in the vicinity of $\mathbf{x}_2^{\pm}$ in physical space.]{\includegraphics[width=0.5\columnwidth]{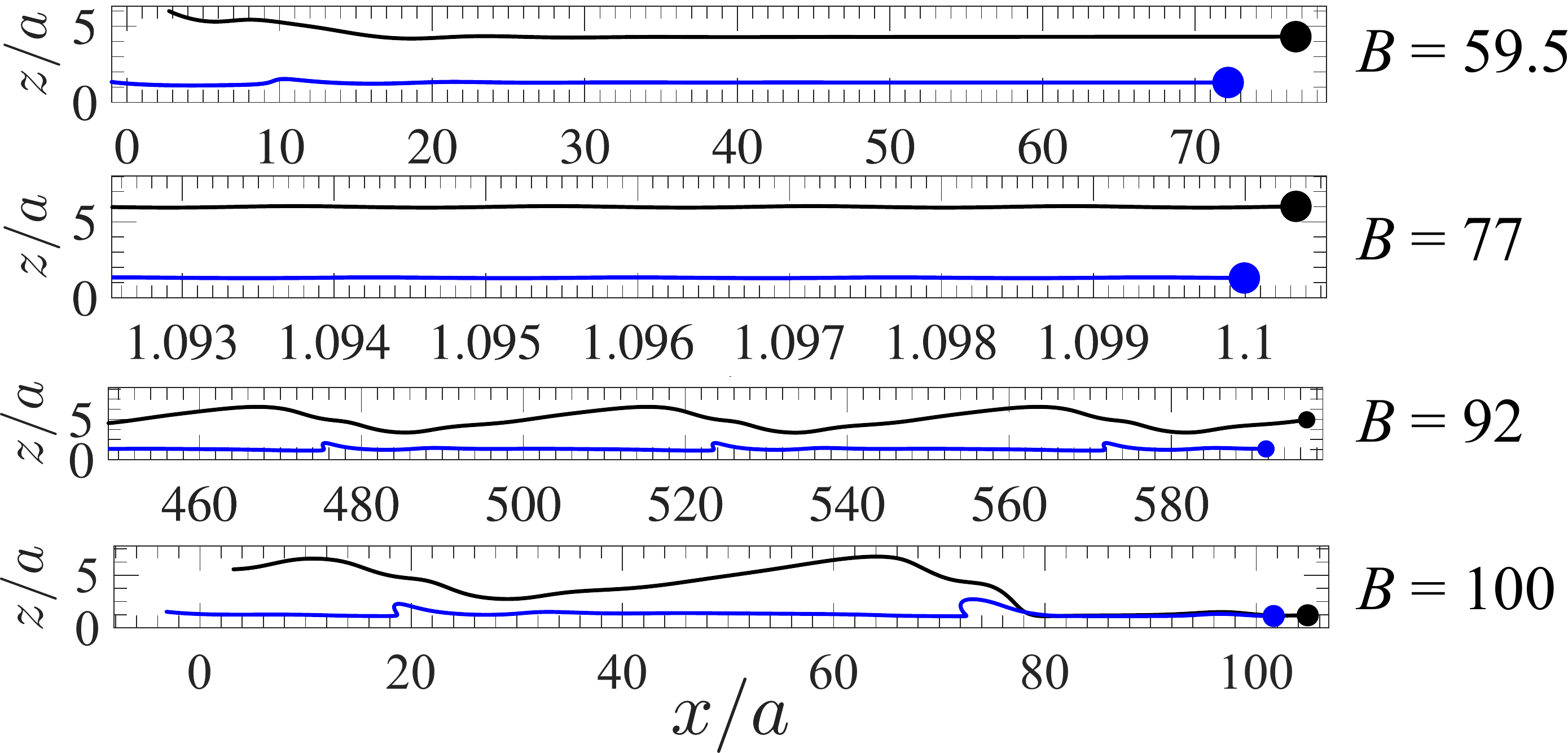}}
    \caption{ Characterization of the Hopf bifurcation at the fixed points $\mathbf{x}^{\pm}_2$} 
    \label{fig:B_phase_space_fixed_point_2}
\end{figure}

When $B=59.5$, two additional symmetric spiral sinks, $\mathbf{x}_2^{\pm}$, appear in the system. In this situation, the active flows are strong enough to counterbalance gravity and maintain a stable configuration where one particle, at the front, is lifted by the other one that is closer to the floor (see Figure \ref{fig:sketch_microrollers}b and first panel in Figure \ref{fig:B_phase_space_fixed_point_2}c). But as $B$ increases, gravity is weaker and loses its role of stabilizer, which leads to the appearance of a limit cycle at $B=77$. As $B$ increases further, the limit cycle grows and disappears for $B>95.5$ (see Figure \ref{fig:B_phase_space_fixed_point_2}c). 
%Such growth and then destabilization of the limit cycle is likely the result of a homoclinic bifurcation. 
We carry out a linear stability analysis of the system in the vicinity of  $\mathbf{x}_2^{\pm}$.
The eigenvalues of $\nabla \mathcal{G}(\mathbf{x}_2^{\pm})$, plotted in Figure \ref{fig:B_phase_space_fixed_point_2}a, show that the linearized system undergoes a Hopf bifurcation at $B=77$: the real part of the complex conjugate eigenvalues becomes positive, which corresponds to the birth of a limit cycle in the nonlinear case.  Linearized trajectories in the vicinity of $\mathbf{x}_2^{\pm}$ are shown in Figure  \ref{fig:B_phase_space_fixed_point_2}b. Contrary to $\mathbf{x}_1^{\pm}$, the coordinates of $\mathbf{x}_2^{\pm}$ change with $B$. In particular the height of the center of mass $z_C$ increases from $2.68a$ to $3.92a$, and the vertical separation $z_{12}$ increases from $2.75a$ to $5.17a$. As illustrated by the first and second panel of Figure \ref{fig:B_phase_space_fixed_point_2}c, when $B$ increases, active flows become stronger compared to gravity and therefore increase the equilibrium height of the front particle, while the rear particle remains approximately at a constant height of $1.3a$.

It is important to mention that these trajectories are purely hydrodynamic in origin. The particles never touch each other nor the wall, thus our results do not depend on the details of the contact forces.  

\section{Conclusions and discussion}
\label{sec:conc_disc}
We have analyzed the phase space of a simple  dynamical system: two coplanar microrollers above a no-slip boundary. We use this elementary model as the starting point to identify the conditions for the spontaneous self-assembly of microrollers into stable motile structures. 
Despite its apparent simplicity, we find that this system exhibits a wide variety of complex behaviors that are controlled by a dimensionless parameter $B$, which compares the external torque to gravity. 
In the limit $B\rightarrow \infty$, active particles are neutrally buoyant and the phase space is divided into two regions: a region with neutrally stable periodic leapfrog trajectories, and a region of diverging  trajectories. Leapfrog trajectories can only be obtained when particle reach the same height with a separation distance below a threshold value $x_{12}^*$. 

For finite values of $B$,  the system exhibits various attractors whose existence and stability depend on $B$. In particular, the leapfrog motion only exists above a critical value $B^*$, for which active flows can overcome gravity. This leapfrog motion is a unique limit cycle whose basin of attraction increases with $B$. Even though our hydrodynamic interactions are not fully resolved and  overestimate particle mobilities, we conjecture that more resolved solutions would not affect the existence of this threshold value, but only shift $B^*$ to a higher value.
Leapfrogging motion of pairs of particles has been observed in previous experiments with dilute suspensions of microrollers \cite{Delmotte2017}, and is thought to be at the origin of the self-assembly into critters. 

We have also discovered the existence of an another stable state at $B=59.5$, where a particle at the rear lifts  another particle up at the front.  This fixed point undergoes a Hopf bifurcation at $B=77$ and then disappears at $B=95.5$.
The existence of this fixed point is also of importance since, before self-assembling into critters, microrollers at the front are lifted up the floor by the others at the rear.\\
%While the existence of a Hopf bifurcation in the system is interesting in itself, the restricted domain of existence ($B \in [59.5;95.5]$) of this fixed point limits its use for particle transport.

Given its simplicity and rich behavior, this system could be used  as an introduction to Dynamical Systems and Fluid Dynamics at the (under-)graduate level. The visual nature of the results and their strong connection to experiments, simulations and current state-of-the-art research, provide a potentially useful tool for science outreach.\\

So far we have only focused on the limit of  large P\'eclet number (Pe$\rightarrow \infty$), where thermal diffusion can be neglected compared to convection. At finite Pe, two dimensionless numbers can be defined in the system, the P\'eclet number: Pe $=Ua/D = U\eta a^2/k_BT = \tau/k_BT$ where $U=\tau/\eta a^2$ is a typical advective velocity induced by the external torque, $k_B$ is the Boltzmann constant, $T$ the solvent temperature;  and the gravitational height $h_g$ that balances gravity with thermal diffusion: $h_g/a = kT/mga$ (note that Pe$\times h_g/a = \tau/mga = B$). At finite Pe, the  motion of the microrollers is given by the overdamped Langevin equations, which require specific time-integration scheme \cite{Delong2014,Delong2015,Delmotte2015,Sprinkle2017}. Using our computational tools \cite{Usabiaga2016b}, we will study the effect of Pe and $h_g$ on the trajectories in phase space in the near future. In particular, it will be interesting to evaluate the robustness of the fixed points and limit cycles versus noise.

In this work we have chosen to focus only on two coplanar particles for two reasons. First, translational invariance permits a description of the system with only three degrees of freedom. The three-dimensional phase space can therefore be visualized and analyzed easily. Second, we are mostly interested in the height dynamics of the system and not in the transverse motion, that we have already analyzed in previous work \cite{Driscoll2017,Delmotte2017b}.
Considering transverse motion along the $y$-direction,  adds one more equation in Eq. \ref{eq:reduced1} and adds one more dimension to the phase space. 

In our large scale simulations and experiments \cite{Driscoll2017}, the critters are made of hundreds or thousands of microrollers. 
The number of degrees of freedom, i.e.\ the dimension of the phase space, for $N$ particles in the $xz$-plane is $N(N-1)+1$.  Detecting and identifying all the fixed points and limit cycles in such high dimensions is not tractable nor is it useful. Instead of adding more and more particles to our dynamical system \eqref{eq:reduced1}, we plan to study microroller suspensions and self-assembly in the continuum limit. We will rely on the formalism developed in \cite{Delmotte2017} to derive a conservation equation for the number density of microrollers, where the advective fluxes are evaluated using convolution integrals with a hydrodynamic interaction kernel and the number density everywhere in the domain.  These hydrodynamic interaction kernels are directly obtained from the mobility functions  \cite{Blake1974,Rotne1969,Yamakawa1970,Swan2007}, and  therefore account for the finite particle size.  Solving the conservation equation numerically (e.g.\ using a high order finite-volume scheme),  we will  determine the influence of $B$ on  the dynamics and on the steady-state structures of the system in two and three dimensions. Once a steady state structure is found, its stability will be analyzed numerically and analytically. 
As shown by our previous work \cite{Driscoll2017}, critters can be used for guided particle transport, flow generation and mixing in microfluidic systems.
We will use these theoretical results together with large scale numerical simulations to optimize the critters generated in experiments. 

\section*{Acknowledgement}
I thank Michelle Driscoll and Aleksandar Donev for their critical reading of the manuscript and Paul Chaikin and Daniel Abrams for insightful discussions on this work.
This work was supported primarily by the Materials Research Science and Engineering Center (MRSEC) program of the National Science Foundation under Award Number DMR- 1420073.  Additional support was provided by the Division of Chemical, Bioengineering, Environmental and Transport Systems program of the National Science Foundation under award CBET-1706562.

\bibliographystyle{unsrt}
\bibliography{Dynamical_system_microrollers_bib}

\end{document}